\documentclass[superscriptaddress,eqsecnum,secnumarabic,preprint,nofootinbib]{revtex4}
\usepackage[mathscr]{eucal}
\usepackage{amscd}
\usepackage{amssymb,latexsym}
\usepackage{verbatim}

\usepackage{amsmath}
\usepackage{amsthm}
\usepackage{enumerate}

\newtheorem{thm}{Theorem}[section]

\newtheorem{prop}[thm]{Proposition}

\newcommand\calV{{\mathcal{V}}}

\newcommand\calT{{\mathcal{T}}}

\renewcommand\l{\lambda}

\newcommand\bbR{{\mathbb R}}

\renewcommand\S{\Sigma}

\renewcommand\d{\partial}

\newcommand\f{\phi}

\newcommand\D{\nabla}
\newcommand\e{\epsilon}

\renewcommand\div{{\rm div}}

\newcommand\la{\langle}
\newcommand\ra{\rangle}

\renewcommand\l{\lambda}
\newcommand\g{\gamma}

\renewcommand\th{\theta}

\newcommand\<{\la}
\renewcommand\>{\ra}

\newcommand\beq{\begin{equation}}
\newcommand\eeq{\end{equation}}
\newcommand\ben{\begin{enumerate}}
\newcommand\een{\end{enumerate}}
\newcommand\bit{\begin{itemize}}
\newcommand\eit{\end{itemize}}

\newcounter{mnotecount}[section]

\begin{document}

\title{Nonexistence of marginally trapped surfaces and geons in $2+1$ gravity}

\author{Gregory J. Galloway}
\affiliation{Department of Mathematics, University of Miami, Coral Gables FL 33158, 
U.S.A.}
\author{Kristin Schleich}
\author{Donald M. Witt}
\affiliation{Department of Physics and Astronomy, University of British Columbia,
Vancouver, British Columbia, Canada,  V6T 1Z1}
\affiliation{Perimeter Institute for Theoretical Physics, 31 Caroline Street North, Waterloo,
Ontario, Canada N2L 2Y5}

\vspace{.2in}

\begin{abstract}  
We use  existence results for Jang's equation and marginally outer trapped surfaces (MOTSs) in $2+1$ gravity  to obtain nonexistence of geons in $2+1$ gravity. In particular, our results show that any $2+1$
initial data set, which obeys the dominant energy condition with cosmological constant $\Lambda \geq 0$ and which satisfies a mild asymptotic condition,
must have trivial topology.  Moreover, any data set obeying these conditions cannot contain a MOTS. 
The asymptotic condition involves a cutoff at a finite boundary at which a null mean convexity condition is assumed to hold; this null mean convexity condition is satisfied by all the standard
asymptotic boundary conditions. 
The results presented here strengthen various aspects of previous related results in the literature.    These results not only have implications for classical $2+1$ gravity but also apply to quantum $2+1$ gravity  when formulated using Witten's solution space quantization.
\end{abstract}

\maketitle

\section{Introduction}  

 Solitons, an interesting feature of  many nonlinear field theories, are  stable solutions that exhibit the characteristics of particles, including properties such as mass, charge and spin. When present, they interact with other particles and fields in the nonlinear theory with important physical consequences.   In gravity, the existence of such solutions, termed geons, was first proposed by Wheeler  in both classical and quantum contexts \cite{Wheeler:1955zz}.  In the original framework,  geons are asymptotically flat solutions of 
Einstein-Maxwell theory. Initial investigations into their existence and properties were carried out in a series of papers by Wheeler and collaborators \cite{Wheeler:1957mu,Misner:1957mt,Brill:1957,Ernst:1957,Brill:1964}. It was
 discovered that geons with trivial topology were  classically unstable on short timescales. In contrast, topological geons do not disperse classically as their nontrivial spatial topology is preserved by evolution under the Einstein equations. Their nontrivial topology also can produce electric charge without the presence of charged matter sources; however, simple types of topological geons, for example those with the topology of a handle, also produce magnetic charge, in contradiction to observed properties of matter coupled to electromagnetism.

An explanation resolving this contradiction and other novel results
 led to renewed interest in topological geons as quantum particles in $3+1$-dimensional quantum gravity. 
Sorkin demonstrated that the nonorientable handle produced electric charges without also producing magnetic monopoles  \cite{Sorkin:1979ja}. Additionally, an interesting formal argument in $3+1$-dimensional quantum gravity demonstrated that  certain topological geons produce spin 1/2 quantum states even though no fermionic matter sources are included
\cite{Friedman:1980st,Friedman:1982du,Sorkin:1985bg}. 
A detailed analysis of the formal 
existence of spin 1/2 states from quantum geons yielded interesting ties to the topology of 3-manifolds, as described in the series of 
papers \cite{Friedman:1983ft, Witt:1986ef, Friedman:1986ze, Friedman:1988he}.\footnote{These 
results also yielded counter-examples to some conjectures in $3$-dimensional topology.} 
Furthermore, physically reasonable initial data sets for the Einstein equations 
can be constructed on all smooth 3-manifolds \cite{Witt:1986ng}; consequently,  
 classical topological geons exist in $3+1$-dimensional gravity. Thus, by the correspondence principle, so should their quantum counterparts in a theory of  $3+1$-dimensional quantum gravity. 

Though intriguing,  these formal arguments regarding the properties of topological geons cannot be more rigorously developed in a quantum context as no complete theory of $3+1$-dimensional quantum gravity is known. However, the potential for such studies exists in one lower dimension; as shown by Witten using a solution space quantization, $2+1$-dimensional quantum gravity  is a well defined theory \cite{Witten:1988hc}. Though initial work concentrated on its formulation for spatially closed $2$-manifolds  \cite{Witten:1988hc, Carlip:1998uc,Carlip:2004ba}, more recent investigations have been in the context of  $2+1$-dimensional anti-de Sitter 
spacetimes \cite{Witten:2007kt} and related $2+1$-dimensional theories with asymptotic regions such as topologically massive gravity \cite{Carlip:2008jk,Carlip:2008eq} and chiral gravity \cite{Li:2008dq}. Consequently, $2+1$-dimensional quantum gravity may provide a natural testbed for rigorously exploring the quantum properties of topological geons.

A natural  first step toward the study of quantum geons in $2+1$-dimensional gravity is the identification of classical $2+1$-dimensional geons.  This paper will rigorously address this issue; are there classical topological geons 
in $2+1$ gravity? 
This question was recently considered for asymptotically flat spacetimes obeying the dominant energy condition in \cite{Stevens:2008hv}. They proved the nonexistence of asymptotically flat  geons  in $2+1$-dimensional vacuum spacetimes and under the more general assumption that
spacetime is analytic.\footnote{Analyticity is used to handle the case of equality in the dominant energy condition. As pointed out in \cite{Stevens:2008hv},  in $2+1$ dimensions, analyticity necessarily holds  for vacuum
spacetimes. }    It follows that there are no quantum geons in its corresponding solution space quantization.  The proof of nonexistence of geons given in \cite{Stevens:2008hv} is based on a spacetime approach that makes use of  topological 
censorship techniques \cite{Friedman:1993ty,Galloway:1999bp,Galloway:1999br}, combined with a refinement of the marginally trapped surface results in $2+1$ gravity considered in  \cite{Ida:2000jh}. 

The aim of the present work is to strengthen various aspects of the nonexistence result  obtained in  \cite{Stevens:2008hv}, which, in the process, involves improvements of  results of  \cite{Ida:2000jh}.  Here we take an initial data set approach, and hence our results
are localized in time.   Moreover, we are able to remove the analyticity assumption in  \cite{Stevens:2008hv};  smooth (or sufficiently differentiable, $C^2$, say) initial data sets suffice.  Also, in  \cite{Stevens:2008hv} implicit
assumptions were made about the existence of outermost marginally outer trapped surfaces (outermost MOTSs).  Here we make careful use of recently established existence results for outermost 
MOTSs \cite{AM1, AM2, Eichmair1, Eichmair2, AEM}. 

The main result of the paper (Theorem \ref{main}) is presented in Section \ref{sec:main}.  In it, we prove that  bounded domains, satisfying a mild and physically natural boundary convexity condition, in $2+1$-dimensional initial data sets obeying the dominant energy condition, with cosmological constant $\Lambda \geq 0$, are necessarily topological disks  and do not contain MOTSs.  (In this work, as will be seen, the cosmological constant is not considered as a source.) 
The proof  makes use of Jang's equation with a Dirichlet boundary condition (as in \cite{SYbhole, Yau}), together with various results about  MOTSs. The advantage of using the Dirichlet boundary condition is that no asymptotic fall-off conditions are needed; the boundary convexity condition mentioned above suffices.   

In Section \ref{sec:mots}, we present some background material on MOTSs and obtain a
strengthening of the results on trapped surfaces in $2+1$ gravity given in \cite{Ida:2000jh}.
This allows  the weakening of the regularity condition used in \cite{Stevens:2008hv}; see especially, 
Theorem~\ref{rigid}, which extends the main rigidity result obtained in \cite{Gal}.   Background material and relevant results on Jang's  equation are presented in Section \ref{sec:jang}.   We emphasize
the connection between MOTSs and Jang's equation to obtain the so-called Schoen-Yau stability
inequality, which plays a key role in the proof of Theorem \ref{main}.

While our results rule out the existence of MOTSs and nontrivial topology in $2+1$-dimensional asymptotically flat initial
data sets obeying the dominant energy condition with $\Lambda \ge 0$, they do not do so for $\Lambda<0$.  Indeed, there are well-known examples of $2+1$-dimensional asymptotically AdS spacetimes which have MOTSs and nontrivial topology, such as the BTZ black holes and related spacetimes 
\cite{Banados:1992wn,Brill:1995jv,Aminneborg:1997pz,Aminneborg:1998si}.   Hence the study of the quantum properties of $2+1$-geons in asymptotically AdS spacetimes remains an intriguing possibility.

{\bf Acknowledgements.} This work was supported by NSF grant DMS-0708048 (GJG) and NSERC Discovery Grants (KS) and (DMW). 
In addition, the authors KS and DMW would like to thank the Perimeter Institute for its hospitality during the period when the ideas 
for this paper were inspired. The authors would also like to thank Michael Eichmair and Jan Metzger for many helpful comments.

\section{Marginally trapped surfaces}
\label{sec:mots}

Let $\S$ be a co-dimension two  spacelike submanifold 
of a spacetime $M$.  Under suitable orientation assumptions, there exist two families
of future directed null geodesics issuing orthogonally from $\S$.  If one of the families has vanishing expansion 
along $\S$ then $\S$ is called a marginally outer trapped surface.  The notion of a
marginally outer trapped surface  (MOTS) was introduced early on in the development of the
theory of black holes \cite{HE}.  Under suitable circumstances, the occurrence of a MOTS signals the presence of a black hole \cite{HE, CGS}.   For this and other reasons MOTSs have played  a fundamental role in quasi-local descriptions of  black holes; see e.g.,  \cite{AK}.  MOTSs arose in a more purely mathematical context 
in the work of Schoen and Yau \cite{SY2} concerning the existence
of solutions to Jang's equation (see Section \ref{sec:jang}), in connection with their proof
of positivity of mass.   

In the following subsections we give precise definitions and present some results about MOTSs relevant to the present work.

\subsection{MOTSs in initial data sets}

In this paper we are primarily interested in initial data sets, and MOTSs therein.

Let $(M^{n+1}, g)$ denote a  spacetime, by which we mean a smooth (Hausdorff, paracompact) manifold $M$ of dimension $n+1$, $n \ge 2$, equipped with a metric $g$ of Lorentz signature $(-+\cdots+)$, such that, with respect to $g$, $M$ is time oriented. 
An initial data set in $(M^{n+1}, g)$ is a triple $(V^n, h, K)$, where $V$ is a spacelike
hypersurface in $M$, and $h$ and $K$ are the induced metric and second fundamental
form, respectively, of $V$.  To set sign conventions, for vectors $X,Y \in T_pV$, $K$ is defined as, $K(X,Y) = \<\D_X u,Y\>$, where $\D$ is the Levi-Civita connection of $M$ and $u$ is the future directed timelike unit vector field to $V$.  Note that a triple $(V^n, h, K)$,
where $V$ is a smooth manifold, $h$ is a Riemannian metric on $V$, and $K$ is a covariant symmetric $2$-tensor on $V$, is always the initial data set of some spacetime (e.g., let   $M'$ be a sufficiently small neighborhood of $\{0\} \times V$ in $\bbR \times V$, equipped with the metric,
$g' = -dt^2 + h_t$, where $h_t = h +tK$).  However, we will only be interested in {\it physically relevant} initial data sets, i.e., initial data sets associated with spacetimes that satisfy the Einstein equations (see Section \ref{sec:variation}).

Let $(V^n, h, K)$ be an initial data set, and   let $\S^{n-1}$ be a closed (compact without boundary) two-sided hypersurface in $V^n$.  
Then $\S$ admits a smooth unit normal field
$\nu$ in $V$, unique up to sign.  By convention, refer to such a choice as outward pointing. 
Then $l_+ = u+\nu$ (resp. $l_- =  u - \nu$) is a future directed outward (resp., future directed inward) pointing null normal vector field along $\S$, unique up to positive scaling.   

The second fundamental form of $\S$ can be decomposed into two scalar valued {\it null second fundamental forms}  $\chi_+$ and $\chi_-$, associated to $l_+$ and $l_-$, respectively.
For each $p \in \S$, $\chi_{\pm}$ is 
the bilinear form defined by,
\beq
\chi_{\pm} : T_p\S \times T_p\S \to \mathbb R , \qquad \chi_{\pm}(X,Y) = g(\D_Xl_{\pm}, Y) \,.
\eeq
The null expansion scalars  $\th_{\pm}$ of $\S$   are obtained by tracing 
$\chi_{\pm}$ with respect to the induced metric $\g$ on $\S$,
\begin{align}
\theta_{\pm} = {\rm tr}_{\g} \chi_{\pm} = {\g}^{AB}\chi_{\pm AB} = {\rm div}\,_{\S} l_{\pm}  \,.
\end{align}
where $\g$ is the induced metric on $\S$.   It is well known that the sign of $\th_{\pm}$ is invariant under positive scaling 
of the null vector field $l_{\pm}$. Physically, $\th_+$ (resp., $\th_-$) measures the divergence
of the  outgoing (resp., ingoing) light rays emanating
from $\S$.  In terms of the initial data $(V^n,h,K)$,
\beq\label{nullid}
\th_{\pm} = {\rm tr}_{\g} K \pm H \,,
\eeq 
where $H$ is the mean curvature of $\S$ within $V$ (given by the divergence of $\nu$ along~$\S$). 

We say that
$\S$ is an {\it outer trapped surface} (resp., {\it weakly
outer trapped surface}) if $\th_+ < 0$ (resp., $\th_+ \le 0$).
If $\th_+$ vanishes, we say that $\S$ is a {\it marginally
outer trapped surface}, or MOTS for short.    Geometrically, MOTSs may be viewed as spacetime analogues of minimal surfaces in Riemannian manifolds.  In fact, in the time-symmetric case ($K = 0$), a MOTS $\S$ is just a minimal surface in $V$.  In recent
years MOTSs have been shown to share a number of properties in common with minimal surfaces.  In particular MOTSs admit a notion of stability analogous to that of minimal surfaces \cite{AMS1,AMS2}.  Here, stability is associated with variations of the null expansion under  deformations of a MOTS.     

\subsection{Variation of the null expansion}
\label{sec:variation}

Let $(V^n, h, K)$ be an initial data set in a spacetime $(M^{n+1},g)$ that obeys
the Einstein equation with cosmological term,
\beq\label{ein}
{\rm Ric} - \frac12 Rg +\Lambda g  = \calT \,,
\eeq
where $\calT$ is the energy-momentum tensor.    The Gauss-Codazzi equations
imply the Einstein constraint equations, 
\begin{align}
\frac12\left(S_V + ({\rm tr}\,K)^2 - |K|^2\right)  &= \rho + \Lambda\\
 \div K - d({\rm tr}\, K) &= J  \,,
\end{align}
where $\rho = \calT(u,u)$,  $J = \calT(u,\cdot)$, and $S_V$ is the scalar curvature of $V$.
For a given choice of $\Lambda$, $\rho$ and $J$ are completely determined by the initial
data.  

The energy-momentum tensor $\calT$ is said to obey the dominant energy condition (DEC)
provided, $\calT(X,Y) = T_{ij}X^iY^j \ge 0$ for all future directed causal vectors $X$ and~$Y$.
One verifies that $\calT$ obeys the DEC if and only for all initial data sets $(V^n,h,K)$ in
$(M^{n+1},g)$,  $\rho \ge |J|$.

We now want to consider variations in the null expansion due to deformations of a MOTS.
Hawking \cite{HE, Hawking:1973} introduced such variational techniques  to obtain results about the topology
of black holes in $3+1$ dimensions. These results were more recently generalized to higher dimensions \cite{GS,Gal}.  Ida
\cite{Ida:2000jh} adapted Hawking's argument to $2+1$  dimensions to obtain restrictions on the existence of certain 
types of MOTSs.

Let $(V^n, h, K)$, $n \ge 2$, be an initial data set in a spacetime obeying the Einstein equations. Let $\S$ be a connected MOTS in $V$ with outward unit normal $\nu$.   We consider
variations $t \to \S_t$   of  $\S = \S_0$, 
$- \epsilon < t  <  \epsilon,$ with variation vector field 
$\calV = \left . \frac{\d}{\d t}\right |_{t=0} = \phi \nu$,  $\phi \in C^{\infty}(\S)$.  Let $\th(t)$ denote
the null expansion of $\S_t$ with respect to $l_t = u + \nu_t$, where $u$ is the future directed timelike unit normal to $V$ and $\nu_t$ is the
outer unit normal  to $\S_t$ in $V$.   A computation \cite{AMS2} shows,
\beq\label{der}
\left . \frac{\d\th}{\d t} \right |_{t=0}   =   L(\f) 
\eeq
where $L : C^{\infty}(\S) \to C^{\infty}(\S)$ is the operator,
\begin{align}\label{op}
L(\phi)  = -\triangle \phi + \<X,\D\phi\>  + 
\left( \frac12 S - P+{\rm div}\, X - |X|^2 \right)\phi  \,,
\end{align}
where,
\begin{align}
S &= 
\begin{cases} 0 \,,  & \text{if $n = 2$}  \\
\text{the scalar curvature of $\S$\,,} & \text{if $n \ge 3$} \,,
\end{cases}   \\
P &=  \rho + J(\nu) + \Lambda  + \frac12 |\chi|^2  \label{p}
\end{align}
($\chi$ = the outward null second fundamental form of $\S$),
and where $X$ is the vector field on $\S$ metrically dual to the one-form, $K(\nu,\cdot)$,  and $\<\cdot,\cdot\> = \g$  is the induced metric  on $\S$.

In the time-symmetric case ($K=0$),  $\th$ becomes the mean curvature $H$, the vector field
$X$ vanishes and $L$ reduces to the classical stability operator of minimal surface theory.
In analogy with the minimal surface case, we refer to $L$ in \eqref{op} as the stability
operator associated with variations in the null expansion $\th$.  Although in general $L$ is not self-adjoint,  its principal eigenvalue (eigenvalue with smallest real part) 
 $\l_1(L)$ is real.   Moreover there exists an associated eigenfunction $\phi \in C^{\infty}(\S)$ which is strictly positive. 
 
As an application of the variational formula (\ref{der}-\ref{op}), we consider the following result, which summarizes several results in the literature \cite{Hawking:1973, Ida:2000jh, GS, Oz}.

\begin{thm}\label{trapped}
Let $(V^n,h,K)$, $n \ge 2$, be an initial data set in a spacetime satisfying the Einstein equations, with  $\Lambda \ge 0$. 
Let $\S$ be a connected MOTS in $V$ such that either (1) $\Lambda = 0$,  and $\rho > |J|$  
along $\S$, or (2)  $\Lambda > 0$, and $\rho \ge  |J|$ along $\S$.  Suppose, further, that
one of the following conditions holds.

\ben[(i)]
\item $n = 2$.
\item $n \ge 3$ and $\int_{\S} Sd\mu  \le 0$.
\item $n \ge 3$ and $\S$ is not of positive Yamabe type, i.e., $\S$ does not admit a metric of positive scalar curvature.
\een 
Then $\S$ can be deformed outward to a strictly outer trapped surface. 
\end{thm}

Ida's \cite{Ida:2000jh} main observation is the case $n =2$.     
Note that in this case $\S$ is one-dimensional and hence is topologically a circle.

\proof  We present here a fairly uniform proof of Theorem \ref{trapped}, which is relevant to the proof of Theorem \ref{rigid} below.   Note that, by the energy conditions,  the scalar quantity $P$ in \eqref{op} is strictly positive.

Consider the ``symmetrized" operator
$L_0: C^{\infty}(\S) \to C^{\infty}(\S)$,
\begin{align}\label{symop}
L_0(\phi)  = -\triangle \phi  + ( \tfrac12 S -P )\phi \,,
\end{align}
obtained from \eqref{op} by formally  setting $X= 0$.   The main argument in \cite{GS}
establishes the following (see also \cite{AMS2}, \cite{Gal}).

\begin{prop}\label{eigen}  
$\l_1(L) \le \l_1(L_0)$.
\end{prop}

For self-adjoint operators of the form \eqref{symop}, the Rayleigh formula \cite{Evans} and an integration by parts give the following standard characterization of the principle eigenvalue,
\beq\label{ray}
\l_1(L_0) = \inf_{\phi \not\equiv 0} \frac{\int_{\S} |\D\phi|^2 + (\tfrac12 S - P)\phi^2 \,d\mu}{\int_{\S} \phi^2 \,d\mu}  \,.
\eeq
In the cases (i) and (ii), we have $\int_{\S} S d\mu \le 0$.  Hence, by setting $\phi = 1$ in the expression on the right hand side of  \eqref{ray},
and using the fact that $P > 0$, we see that $\l_1(L_0) < 0$.   Thus, by Proposition
\ref{eigen}, $\l_1(L) < 0$.   

Now let $\phi$ be an eigenfunction associated to $\l_1(L)$,
$L(\phi) = \l_1(L) \phi$; $\phi$ can be chosen to be strictly positive.
 Using this $\phi$ to
define our variation $t \to \S_t$, we have from (\ref{der}),
\beq\label{der2}
\left . \frac{\d\th}{\d t} \right |_{t=0} =\l_1(L) \phi  < 0 \,.
\eeq
Together with the fact that $\th = 0$ on $\S$, this implies 
that for $t>0$ sufficiently small, $\S_t$ is outer trapped, as desired.

Now consider case (iii).  First suppose $n = 3$. Then $\S$ is $2$-dimensional, and by the Gauss-Bonnet theorem, the assumption that $\S$ does not carry a metric of positive curvature implies $\int_{\S} S d\mu \le 0$.  Thus, the argument is the same as in cases (i) and (ii).  For $n \ge 4$, 
consider the conformal Laplacian,  
$L_{cf} : C^{\infty}(\S^{n-1}) \to C^{\infty}(\S^{n-1})$,
\beq
L_{cf} (\phi) = -4\frac{n-2}{n-3} \triangle\phi + S\phi \,.
\eeq
If $\S$ does not carry a metric of positive scalar curvature then we must have, 
$\l_1(L_{cf}) \le~0$ \cite{Besse} .  The Rayleigh formula applied to $L_{cf}$ gives,
\beq\label{ray2}
\l_1(L_{cf}) = \inf_{\phi \not\equiv 0}  \frac{\int_{\S} \tfrac{4(n-2)}{n-3}|\D\f|^2 +S\f^2\, d\mu}
{\int_{\S} \phi^2 \,d\mu} \, .
\eeq
Comparing \eqref{ray} and \eqref{ray2}, and using the positivity of $P$, one easily obtains,
$\l_1(L_0) < \frac12 \l_1(L_{cf})$.  Hence, $\l_1(L_0) < 0$, and so by  Proposition \ref{eigen}, we again arrive at, $\l_1(L) < 0$.  We  may then proceed as before.~\qed

\smallskip
One can see, by a simple modification of the proof, that in the case $\Lambda = 0$,
it is sufficient to require $\rho \ge |J|$ on $\S$, with strict inequality somewhere. 

With somewhat more effort, one can obtain the following refinement of Theorem \ref{trapped}
which does not require any strictness in the energy conditions.

\begin{thm}\label{rigid}
Let $(V^n,h,K)$, $n \ge 2$, be an initial data set in a spacetime satisfying the Einstein equations \eqref{ein} with $\Lambda \ge 0$, such that $\calT$ satisfies the DEC.  
Suppose  $\S$ is a connected MOTS in $V$ such that in some neighborhood $U\subset V$ of $\S$ there are
no (strictly) outer trapped surfaces outside of, and homologous, to $\S$.  
Suppose, further, that one of the following conditions holds.
\ben[(i)]
\item $n = 2$.
\item $n = 3$ and $\int_{\S} Sd\mu  \le 0$.
\item $n \ge 3$ and $\S$ is not of positive Yamabe type, i.e., $\S$ does not admit a metric of positive scalar curvature.
\een 
Then there exists an outer half-neighborhood $U^+$ of $\S$  foliated by MOTSs, i.e.,
$U^+ \approx [0,\e) \times \S$,  such that each slice
$\S_t = \{t\} \times \S$, $t \in [0,\e)$ is a MOTS.  
\end{thm}

\proof[Remarks on the proof]  Case (iii) of Theorem \ref{rigid} is proved in \cite{Gal}.  The proof
in this case consists of two steps.  In the first step, one obtains an outer foliation $t \to \S_t$,
$0 \le t \le \e$, of surfaces $\S_t$ of {\it constant} outer null expansion, $\th(t) = c_t$.  The second step involves showing that the constants $c_t = 0$.  This latter step requires a reduction  to the case that $V$ has nonpositive mean curvature, $\tau \le 0$ near $\S$.  For this it is necessary to know that the DEC holds in a spacetime neighborhood of $\S$.  The proof makes use of the formula for the $t$-derivative, $\frac{\d\th}{\d t}$, not just at $t=0$ where
$\th =0$, but all along the foliation $t \to \S_t$, where, a priori, $\th(t)$ need not be zero.  Thus,  additional terms appear in the expression for $\frac{\d\th}{\d t}$ beyond  those in \eqref{der}, including a term involving the mean curvature of $V$, which need to be accounted for.   The proof of
case (iii) given in \cite{Gal} can be easily modified to give a proof of Theorem \ref{rigid}
in the cases (i) and (ii), by using arguments like those used in the proof of the cases
(i) and (ii) in Theorem \ref{trapped} above.  For Theorem \ref{rigid}, it is necessary
to restrict the dimension in case (ii) to $n =3$ in order to control, via Gauss-Bonnet,
the total scalar curvature of each $\S_t$. ~\qed

\smallskip
Let us say that a connected MOTS $\S$ in an initial data set is {\it locally outermost} if, with respect to some neighborhood of $U \subset V$ of $\S$,  there are no weakly outer trapped surfaces outside of, and homologous, to $\S$ in $U$.   Theorem \ref{rigid}(i) shows that for initial data sets
in $2+1$-dimensional spacetimes satisfying the Einstein equations \eqref{ein} with 
$\Lambda \ge 0$, such that $\calT$ satisfies the DEC, there can be no locally outermost MOTSs.  This strengthens Ida's results and those of  \cite{Stevens:2008hv} by removing any strictness in the energy inequalities, restriction to the vacuum case or assumption of analyticity. 
Theorem \ref{rigid}(i) rules out, in particular,  the existence of $2+1$-dimensional stationary black hole spacetimes   obeying the stated energy conditions.  

In Section \ref{sec:main}, we obtain a more comprehensive
result which rules out MOTSs altogether, locally outermost or otherwise. 

\subsection{Existence of MOTSs}

Substantial progress has been made in recent years concerning the existence 
of MOTSs.   Following an approach suggested by Schoen \cite{Schoen},
Andersson and Metzger \cite{AM2}
established for $3$-dimensional initial data sets the existence of MOTSs under natural 
barrier conditions.  Combining this existence result with the compactness
result  established in \cite{AM1} and an interesting surgery technique, they were able to establish the existence of outermost MOTSs
in $3$-dimensional initial data sets \cite{AM2}.   Such an outermost MOTS was realized as the boundary of the so-called {\it trapped region}, suitably defined.   Using insights from geometric measure theory,  Eichmair \cite{Eichmair1,Eichmair2} was able to extend the results of Andersson and Metzger to dimensions $n$, $2 \le n \le 7$.   We refer the reader to the recent survey article~\cite{AEM} for an excellent discussion of these existence results.  

The results concerning the existence of outermost MOTSs (see \cite[Theorem 4.6]{AEM})
may be formulated as follows.

\begin{thm}\label{exist}  Let $(V^n, h, K)$ be an initial data set, $2 \le n \le 7$, and let $W^n$ be a connected compact $n$-manifold-with-boundary in $V^n$.   Suppose that
the boundary $\d W$ can be expressed as a disjoint union, $\d W = \S_{inn} \cup \S_{out}$,
such that $\th^+ < 0$ along $\S_{inn}$ with respect to the null normal whose projection  points into $W$, and $\th^+ > 0$  along $\S_{out}$ with respect to the null normal whose projection  points out of $W$.  Then there
exists a smooth compact outermost MOTS $\S$ in the interior of $W$ homologous to $\S_{out}$.
\end{thm}

Some remarks are in order.
\ben
\item If, as the notation suggests, we think of $\S_{inn}$ as an {\it  inner} boundary and
$\S_{out}$ as an {\it outer} boundary, then we are assuming that $\S_{inn}$
is outer trapped and $\S_{out}$ is outer untrapped.
\item By $\S$ being {\it homologous} to $\S_{out}$, we mean explicitly  that there exists an open set $U \subset W$ such that $\d U = \S \cup  \S_{out}$.  Then $\th_+$ is defined with respect to the null normal whose projection points into $U$.   
\item By $\S$ being {\it outermost} in $W$ we mean that if $\S'$ is a  weakly outer trapped 
($\th^+ \le 0$) surface in $\overline U$  homologous to $\S_{out}$ then $\S' = \S$.  In other words,
$\S$ must enclose all weakly outer trapped surfaces in $W$ homologous to $\S_{out}$.
\item It is important to note for applications that $\S_{inn}$ and $\S_{out}$ need not  be connected.  Also the MOTS $\S$ will not in general be connected (even if $\S_{inn}$ and $\S_{out}$ are).
\item Finally,  Andersson and Metzger \cite[Section 5]{AM2} have shown, by a technique of modifying the initial data near the inner boundary to get a strict barrier, that it is sufficient in  Theorem \ref{exist}  to require that $\S_{inn}$ be only weakly  outer trapped, $\th_+ \le 0$.
Then the outermost MOTS $\S$ may have components that agree with components of 
$\S_{inn}$.
 \een

Note the tension between Theorem \ref{rigid}(i) and Theorem \ref{exist}, the former implying that there are no locally outermost MOTSs under appropriate energy conditions, and the latter providing conditions for the existence of outermost MOTSs.     
We will exploit this tension in the proof of the main result in Section \ref{sec:main}.

The proof of the basic existence result for MOTSs alluded to at the beginning of this subsection  
is based on Jang's equation \cite{Jang}, which we discuss in the next section. 
Schoen and Yau \cite{SY2} established existence and
regularity for Jang's equation with respect to asymptotically
flat initial sets, as part of their approach to proving the
positive mass theorem for general, nonmaximal, initial data
sets.   In the process they discovered an obstruction to global
existence: Solutions to Jang's equation  tend to blow-up in the
presence of MOTSs in the initial data $(V^n,h,K)$.   Turning the situation around, this behavior was exploited in \cite{AM2, Eichmair2} to establish the existence of MOTSs by {\it inducing}
blow-up of Jang's; see \cite{AEM} for further discussion. 

\section{Jang's equation and the Schoen-Yau stability inequality}\label{sec:jang}

Let $(V^n,h,K)$ be an initial data set.  Then Jang's equation is the equation,
\begin{equation} \label{jang}
{\g}^{ij} \left(\frac{D_i D_j f} {\sqrt{1 + |D f|^2}} - K_{ij} \right) = 0 \,,
\end{equation}
where $f$ is a function on $V$, $D$ is the Levi-Civita connection of $h$, and  ${\g}^{ij}=h^{ij} - \frac{f^i f^j}{1 + |D f|^2}$.  Introducing  the Riemannian product manifold, $\bar V = V \times \bbR$, $\bar h = h + dz^2$ , we notice that the $\g^{ij}$'s are the contravariant components of the induced metric
$\g_f$ on $\S_f = {\rm  graph}\,f$ in $\bar V$, and, moreover, that, 
\begin{equation*} 
H(f) := - \frac{\g^{ij}D_iD_j f} {\sqrt{1 + |D f|^2}} 
\end{equation*}
is the mean curvature of $\S_f$, computed with respect to the {\it upward pointing}\footnote{We note that in \cite{SY2}  the mean curvature of $\S_f$ is considered with respect to the downward pointing normal.  Our choice results in some minor sign differences.} unit normal $\nu$.  Thus, Jang's equation becomes,
\beq
H(f) + {\rm tr}\,_{\g_f} \bar K = 0 \,,
\eeq
where $\bar K$ is the pullback, via
projection along the $z$-factor, of  $K$ to $\bar V$.
Comparing with Equation \eqref{nullid}, we see that, geometrically,  Jang's equation is  the requirement that the graph $\S_f$  has vanishing null expansion, $\th_+ =0$, i.e., is a MOTS, in the initial data set $(\bar V^n, \bar h, \bar K)$.

Given a solution $f$  to Jang's equation, we can use Equations (\ref{der}-\ref{op}) to obtain a formula for the scalar curvature $S_f$ of $\S_f$.  Consider the variation $t \to \S(t)$ of $\S_f$
obtained by shifting $\S_f$ up and down the $z$-axis, i.e., $\S(t) =$ the graph of $f+t$. This may be viewed as a normal variation, with variation vector field,
\beq
\calV =   \phi \nu \,, \quad \phi = \bar h(\nu,\d_z) \,
\eeq
where $\nu$ is the upward pointing unit normal along $\S_f$.

Let $\th(t)$ denote the null expansion of $\S(t)$.  Because Jang's equation is translation invariant, in the sense  that if $f$ is a solution then $f+t$ is also a solution, we have that $\th(t) = 0$ for all $t$.  Hence, $\left . \frac{\d\th}{\d t}\right |_{t=0} = 0$, and Equations (\ref{der}-\ref{op}) give
along $\S_f$,
\beq\label{stable}
 -\triangle \phi + \<\bar X,\D\phi\>  + \left( \frac12 S_f - P+{\rm div}\, \bar X - |\bar X|^2 \right)\phi = 0\,,
\eeq
where $\bar X$  is the vector field on $\S_f$ metrically dual to the one-form, $\bar K(\nu,\cdot)$,
and
\beq
P =  \bar \rho + \bar J(\nu) + \Lambda  + \frac12 |\bar\chi|^2  \label{p2}\,,
\eeq
where $\bar \rho$ and $\bar J$ are the pullback of $\rho$ and $J$, respectively, via projection
along the $z$-factor.  

By setting $\phi = e^u$ in \eqref{stable} and completing the square, we obtain,
\beq\label{stabineq}
 \frac12 S_f + \div(\bar X -\D u) - |\bar X - \D u|^2  =    \bar \rho +  \bar J(\nu) + \Lambda  + \frac12 |\bar\chi|^2 \ge 0 \,,
\eeq
where the inequality holds provided $\Lambda \ge 0$ and the DEC, $\rho \ge |J|$, holds
with respect to the original initial data set $(V^n, h, K)$.  This inequality is equivalent to the 
``on shell"  {\it Schoen-Yau stability inequality}\footnote{Here {\it stability} relates to the fact that with respect to the variation being considered, the null expansion is nondecreasing in the ``outward" direction.} obtained in \cite{SY2}; cf., (2.29) on p. 240.
Hence, assuming $\Lambda \ge 0$ and the DEC holds, we arrive at,
\beq\label{scalarineq}
S_f \ge -2\,\div(\bar X -\D u)    \,,
\eeq
where $u = \ln\, \bar h(\nu, \d_z)$ and $\bar X$  is the vector field on $\S_f$ metrically dual to the one-form $\bar K(\nu,\cdot)$.

In \cite{SY2} Schoen and Yau studied extensively the existence and regularity of solutions  $f$ to Jang's equations over complete asymptotically flat $3$-dimensional initial data sets $(V^3, h,K)$, with suitable decay on the asymptotically Euclidean ends.   As noted in~\cite{SYbhole} 
(see also \cite{Yau}), by standard considerations one obtains  similar existence results 
for Jang's equation with Dirichlet boundary data, $f =0$, on compact manifolds $W$ with {\it null convex} boundaries $\d W$ (as defined in the next section).  It follows as  
an immediate consequence of their main existence result  \cite[Proposition 4]{SY2} 
(see also \cite[Theorem 3.2]{AEM}) that  {\it if there are no MOTSs in $W$ then   there exists a globally regular solution $f$ of Jang's equation on $W$ with Dirichlet boundary data $f =0$.} 
This result remains valid for $2$-dimensional initial data sets, and will be used in the proof of the main result.

\section{Main Result}\label{sec:main}

Let $W^n$, $n \ge 2$, be a connected compact manifold-with-boundary in an initial data set $(V^n,h,K)$.  We say that
the boundary $\d W^n$ is {\it null mean convex}
provided it has positive 
outward null expansion, $\th^+ > 0$, and negative inward null expansion, $\th^-< 0$.  
Note that round spheres in Euclidean slices of Minkowski space, and, more generally, large ``radial" spheres in asymptotically flat initial data sets are null mean convex.

The aim of this section is to prove the following result about {\it 2-dimensional} initial data
sets.

\begin{thm}\label{main}
Let $(V^2,h,K)$  be a $2$-dimensional initial data set in a spacetime satisfying the Einstein equations \eqref{ein} with $\Lambda \ge 0$, such that $\calT$ satisfies the DEC.   If $W^2$ is 
a connected compact $2$-manifold with null mean convex boundary $\d W^2$ in $V^2$, then $W^2$ is diffeomorphic to a disk, and there are no MOTSs in $W^2$.
\end{thm}

\noindent
The theorem follows from two claims.

\medskip
\noindent
\underline{Claim 1.} If there are no MOTSs in $W$ then $W$ is diffeomorphic to a disk.

\proof As per the comments at the end of the previous section, if there are no MOTSs in
$W$ then there exists a globally regular solution $f: W \to\bbR$ to Jang's equation, with $f =0$ on $\d W$.  As  in
Section \ref{sec:jang}, we consider $\S_f = $ graph~$f$ in the metric $\g_f$ induced from the 
product metric $\<\cdot,\cdot\> = h + dz^2$.  
We introduce an orthonormal frame ${e_1, e_2, e_3}$ along $\S_f$ near $\d \S_f = \d W$. 
Take $e_3 = \nu$, and let $e_1$ and $e_2$ be tangent to $\S_f$, such that $e_1$ is tangent to $\d\S_f$ and $e_2$ is normal to $\d \S_f$ and outward pointing.

Let $S^2_d$ denote the $2$-sphere with $d \ge 1$ disjoint open disks removed.  By the classification of surfaces, if $W$ is orientable then it  is diffeomorphic to a connected sum of $S^2_d$ and $g$ tori, $g \ge 0$, while if it is nonorientable it is a connected sum of $S^2_d$ and $k$ projective planes, $k \ge 0$.   Then
by the Gauss-Bonnet  formula applied to  $(\S_f,\g_f)$, we have,
\begin{align}\label{gb}
\iint_{\S_f} {\cal K} dA + \int_{\d \S_f} \kappa ds  &= 2\pi \chi(\S_f) = 2\pi\chi(W)  \nonumber\\
& = 2\pi \, (2-a -d) \,.
\end{align}
where $a = 2g$ or $k$, depending on whether $W$ is orientable or nonorientable.

To show that $a = 0$ and $d =1$, and hence that $W$ is a disk, it is sufficient to show that
the left hand side of \eqref{gb} is strictly positive.  From \eqref{scalarineq},
the Gaussian curvature ${\cal K}$ satisfies, ${\cal K} \ge - \div(\bar X - \D u)$, where 
$u = \ln \<e_3, \d_z\>$ and $\bar X$  is the vector field on $\S_f$ metrically dual to the one-form $\bar K(\nu,\cdot)$.
The geodesic curvature
$\kappa$ is given by $\kappa = -\<\D_{e_1}e_1,e_2\> 
= \bar H_{\d W} $, 
the mean curvature of $\d W$ in $(\S_f,\g_f)$ with respect to the outward unit normal $e_2$.
Then, applying the divergence theorem,
\begin{align}\label{gbineq}
\iint_{\S_f} {\cal K} dA + \int_{\d \S_f} \kappa ds & \ge \int_{\d W} \bar H_{\d W} - \<\bar X, e_2\> 
+ \<\D u, e_2\>    \, ds \nonumber \\
& =  \int_{\d W} \bar H_{\d W} -  \bar K(e_3,e_2)  + e_2(u) \, ds  \,.
\end{align}
By analyzing each term in the integrand in a manner similar to what is done in \cite[p. 9f]{Yau},
we show that the integrand is strictly positive.

Let $w$ be the unit normal field to $\d W$ tangent to $V$.  Then note, since $\d_z$ is parallel,
\begin{align}\label{bdry}
 \bar H_{\d W} =  -\<\D_{e_1}e_1,e_2\> = - \<e_2,w\> \<\D_{e_1}e_1,w\> = \<e_2,w\> H_{\d W} \,,
\end{align}
where $H_{\d W}$ is the mean curvature of $\d W$ in $(V, h)$.  Also, since $\bar K(\d_z,\cdot) =0$, 
\begin{align}\label{fundform}
\bar K(e_3,e_2) = \<e_3, w\> \bar K(w,e_2) = \frac{\<e_3, w\>}{\<e_2, w\>} \bar K(e_2,e_2) \,.
\end{align}
For the term  $e_2(u)$, we have,
\begin{align}\label{deru}
e_2(u) &= \frac1{\<e_3, \d_z\>}\, e_2 \<e_3, \d_z\> = \frac1{\<e_3, \d_z\>}\<\D_{e_2}e_3,\d_z\>  \nonumber\\
& = \frac{\<e_2, \d_z\>}{\<e_3, \d_z\>} \<\D_{e_2}e_3,e_2\> = -\frac{\<e_2, \d_z\>}{\<e_3, \d_z\>} \<\D_{e_2}e_2,e_3\> \nonumber \\
& =  \frac{\<e_3, w\>}{\<e_2, w\>} \<\D_{e_2}e_2,e_3\>  \,.
\end{align}
Using the following,
\begin{align}
-{\rm tr}\, \bar K &= H(f) = 
-\<\D_{e_1}e_1, e_3\> -\<\D_{e_2}e_2, e_3\>  = \nonumber\\
& = \<e_3,w\> H_{\d W} -  \<\D_{e_2}e_2, e_3\> \,,
\end{align}
we can write Equation \eqref{deru}  as
\begin{align}\label{deru2}
e_2(u) = \frac{\<e_3, w\>}{\<e_2, w\>} \left(\<e_3,w\> H_{\d W} + {\rm tr}\, \bar K \right)  \,.
\end{align}
Combining \eqref{bdry}, \eqref{fundform}, and \eqref{deru2}, we obtain,
\begin{align}\label{integrand}
\bar H_{\d W} -  \bar K(e_3,e_2)  + e_2(u) &= \left( \<e_2,w\> + \frac{\<e_3, w\>^2}{\<e_2, w\>}\right) H_{\d W} + \frac{\<e_3, w\>}{\<e_2, w\>}\left({\rm tr}\, \bar K - \bar K(e_2,e_2)\right)
 \nonumber    \\
 &= \<e_2,w\>^{-1}\left( H_{\d W}  - \<e_3, w\> \, {\rm tr}_{\d W} K \right)  \nonumber \\
 & \ge \<e_2,w\>^{-1}\left( H_{\d W}  - | {\rm tr}_{\d W} K| \right)  \,.
\end{align}
But  observe that the quantity $H_{\d W}  - | {\rm tr}_{\d W} K|$ is positive if and only if
$\d W$ is null mean convex, and moreover that
$H_{\d W}  - | {\rm tr}_{\d W} K|  \ge \min(\th_+,-\th_-)$. 

Hence, using \eqref{integrand} in  \eqref{gbineq} we obtain,

\begin{align}\label{pos}
\iint_{\S_f} {\cal K} dA + \int_{\d \S_f} \kappa ds & \ge \int_{\d W} H_{\d W}  - | {\rm tr}_{\d W} K| \,ds > 0 \,, 
\end{align}
from which we conclude that $W^2$ is diffeomophic to a disk.\qed

\medskip
\noindent
\underline{Claim 2.} There are no MOTSs in $W$.

\proof  Suppose $\S$ is a (connected) MOTS in $W$.  Then  $\S$ is two-sided and $\th_+ = 0$ with respect to the null normal  $l_+ = u + \nu$, where $\nu$ is a  smooth unit normal to $\S$ 
in~$W$.  

Suppose that $\S$ separates $W$. Then $\S$ is homologous to $\S'$, where 
$\S'$ is a nonempty disjoint union of some (perhaps all) of the components of $\d W$.  That is,  there exists and open set $U \subset W$ with $\d U = \S \cup \S'$. Moreover, by considering the time-dual of spacetime if necessary, we can assume that $\nu$ points  into $U$.  We may now apply Theorem \ref{exist}, together with Remark 5, with $\S_{inn} = \S$ and $\S_{out} = \S'$, to conclude that there exists an outermost MOTS $\hat\S$ in $U \cup \S$.  
On the other hand, by applying Theorem~\ref{rigid}(i) to one of the components of $\hat\S$ we see that $\hat\S$ cannot be  outermost.

Now suppose that $\S$ does not separate $W$.  In this case we modify $W$ by making  a ``cut" along $\S$; as MOTSs are two-sided, this produces a compact surface $W'$ with boundary $\d W' = \d W \cup \S_- \cup \S_+$,
where $\S_-$ and $\S_+$ are copies of $\S$ such that $\S_+$ is a  MOTS with respect to 
the normal pointing into $W'$ and $\S_-$ is a MOTS with respect to the normal pointing out of $W'$.  Now apply Theorem \ref{exist}, together with Remark~5, with  $\S_{inn} = \S_+$ and 
$\S_{out} = \d W \cup \S_-$ to obtain an outermost MOTS in $W'$.\footnote{Strictly speaking to apply Theorem \ref{exist}, one must modify the initial data on an outer tubular neighborhood of $\S_-$  such that this neighborhood is foliated by circles with strictly positive outward null expansion; see \cite[Section 5]{AM2}. }
We note that since $\S_-$ is not homologous to $\S_+$, this outermost MOTS must have
at least one component distinct from $\S_-$.
Applying Theorem \ref{rigid}(i) to this component again leads to a contradiction. 
Thus, there can be no MOTS in $W$.  This completes the proof of Claim 2 and hence Theorem~\ref{main}.\qed


\end{document}